\documentclass[pra, twocolumn, superscriptaddress,  longbibliography,  aps]{revtex4-2}

\usepackage{amsfonts,amssymb,amsmath}

\usepackage{graphicx}
\usepackage[space]{grffile}
\usepackage{latexsym}
\usepackage{textcomp}
\usepackage{longtable}
\usepackage{multirow,booktabs}
\usepackage[english]{babel}
\usepackage{dcolumn}
\usepackage{bm}
\usepackage{units}
\usepackage{upgreek}
\usepackage{color}
\usepackage{cancel}

\newcommand{\be}{\begin{equation}}
\newcommand{\ee}{\end{equation}}
\newcommand{\bea}{\begin{eqnarray}}
\newcommand{\eea}{\end{eqnarray}}

\newcommand{\ket}[1]{|#1\rangle}
\newcommand{\bra}[1]{\left\langle#1\right|}

\newcommand{\fb}{f_\text{CD}}

\newcommand{\sx}{\hat{\sigma}_x}

\newcommand{\sz}{\hat{\sigma}_z}

\DeclareMathOperator{\sign}{sgn}

\begin{document}

\title{Quantum control by effective counterdiabatic driving}

\author{Francesco Petiziol}
\affiliation{Institut f\"ur Theoretische Physik, Technische Universit\"at Berlin, Hardenbergstr. 36, 10623 Berlin, Germany}

\author{Florian Minter}
\affiliation{Department of Physics, Imperial College London, SW7 2AZ London, United Kingdom}
\affiliation{Helmholtz-Zentrum Dresden-Rossendorf, Bautzner Landstrasse 400, 01328 Dresden, Germany}

\author{Sandro Wimberger}
\email{sandromarcel.wimberger@unipr.it}
\affiliation{Dipartimento di Scienze Matematiche, Fisiche e Informatiche, Universit\`{a} di Parma, Parco Area delle Scienze 7/A, 43124 Parma, Italy}
\affiliation{INFN, Sezione di Milano Bicocca, Gruppo Collegato di Parma, Parco Area delle Scienze 7/A, 43124 Parma, Italy}

\begin{abstract}
We review a scheme for the systematic design of quantum control protocols based on shortcuts to adiabaticity in few-level quantum systems. The adiabatic dynamics is accelerated by introducing high-frequency modulations in the control Hamiltonian, which mimic a time-dependent counterdiabatic correction. We present a number of applications for the high-fidelity realization of quantum state transfers and quantum gates based on effective counterdiabatic driving, in platforms ranging from superconducting circuits to Rydberg atoms.
\end{abstract}

\keywords{Counterdiabatic driving, Shortcut to adiabaticity, Floquet engineering, Quantum computing}

\maketitle

\section{Introduction}
\label{sec:intro}

This Perspective reviews a framework to construct quantum control protocols based on accelerated adiabatic evolutions, falling into the class of so-called shortcuts to adiabaticity (STAs)~\cite{Torrontegui2013, Odelin2019, delCampo2019}. A STA attempts to realize the same dynamics obtained by means of a slow adiabatic variation of the control parameters, but in a shorter time. This acceleration is possible, since the STA gives up on following an eigenstate of the instantaneous driven Hamiltonian at all times, and admits temporary excursion into other states instead. 
STAs theoretically guaranteeing exact tracking of the adiabatic states can be constructed based on reverse-engineering procedures, but their direct experimental realization is often challenging due to the necessity to control additional degrees of freedom. The paradigmatic example of an exact STA, which we deal with in the following, is known as transitionless or counterdiabatic (CD) driving. It has its origins in the adiabatic control of atoms and molecules \cite{Berry1987, Lim1991, Unanyan1997}, and its theory was developed in~\cite{Demirplak2003, Berry2009}.

The problem of implementing new controls, not already present in the Hamiltonian describing the adiabatic drive to be sped up, led to the development of approximate CD pulses in many setups \cite{IbanezRuschhaupt2012, opatrny201, Baksic2016, Petiziol2018, Claeys2019, Solano2021, Daley2023, Bukov2023}. The framework reviewed here attempts to overcome this difficulty, by providing a systematic method to approximate a CD Hamiltonian with initially {\em available} controls. It builds on the use of time (quasi-)\ periodic forcing, which is well understood within Floquet theory \cite{Floquet1883, Shirley1965, Salzman1974}. Using periodic drives is one of the most established methods to control atomic \cite{Breuer1989, Casati1989, Doerr1994, GRIFONI1998, BUCHLEITNER2002,  CHU2004, Goldman2014, Holthaus2016, Bukov2015, Eckardt2017, Weitenberg2021}, molecular \cite{Breuer1991, Tannor2007} and spin systems \cite{Andrew1958, MANANGA20161}, for instance. The basic idea of the method reviewed is to create the new Hamiltonian terms in an effective manner \cite{Bukov2015, Eckardt2017, Weitenberg2021}, through a suitable high-frequency modulation of operators already contained in the original adiabatic Hamiltonian.
Experimental realisations of exact or approximate CD driving were shown for Bose-Einstein condensates \cite{BasonMorsch2012}, NV centers \cite{ZhangSuter2013, ZhouAwschalom2017}, ions \cite{An2016, Hu2018}, NMR \cite{Santos2020}, superconducting circuits \cite{Zhang2018, Wang2018, VepsalainenParaoanu:2019, Dogra2022, Ge2023}, or molecules \cite{Zhou2020}. We review the CD method in the next section, and then introduce the effective CD driving in the following one.
After showing a selection of the--by now--many application, we conclude in the last section with an outlook on further advances.

\section{Counterdiabatic driving}
\label{sec:CD}

CD driving builds on a Hamiltonian $\hat{H}_0(\lambda) + \hat{H}_{\mathrm{CD}}(\lambda)$, which generates exactly, in finite time, the same evolution obtained through $\hat{H}_0(\lambda)$ in the limit of an infinitely slow variation of the parameter $\lambda=\lambda(t)$.
Considering the instantaneous energies $E_n(\lambda)$ and eigenstates $\ket{n_\lambda}$, such that $\hat{H}_0(\lambda) \ket{n_\lambda} = E_n(\lambda) \ket{n_\lambda}$, any quantum state can be written as $\ket{\psi{(\lambda)}} = \sum_n a_n(\lambda) \ket{n_\lambda}$.
Schr\"odinger's equation for the expansion coefficients $a_n(\lambda)$ then reads
 \begin{eqnarray}
\partial_\lambda a_n(\lambda) & = & -i \left[\dot{\lambda}^{-1}E_n(\lambda) - i  \langle{n_\lambda | \partial_\lambda n_\lambda}\rangle \right] a_n(\lambda)  \nonumber \\
& & -  \sum_{m \neq n} \frac{\bra{n_\lambda} \partial_\lambda \hat{H}_0(\lambda) \ket{m_\lambda}}{E_m(\lambda)-E_n(\lambda)} a_m(\lambda),
\label{eq:coeffs}
\end{eqnarray}
where we used $\langle n_\lambda|\partial_\lambda m_\lambda\rangle = \bra{n_\lambda}\partial_\lambda \hat{H}_0(\lambda) \ket{m_\lambda}/[E_m(\lambda) - E_n(\lambda)]$ for $n\ne m$. The first line on the r.h.s, including the Berry connection $i\langle n_\lambda |\partial_\lambda n_\lambda \rangle$~\cite{Berry1984}, describes the accumulation of dynamical and geometric phases, while the second line describes nonadiabatic transitions among the instantaneous eigenstates. Such transitions can be exactly canceled by changing the original Hamiltonian to $\hat{H}_0(\lambda) + \hat{H}_{\rm CD}(\lambda)$, with
\begin{equation}
\label{eq:cd-ham}
 \hat{H}_{\rm CD}(\lambda) = i \dot{\lambda} \sum_{n, m \ne n} \frac{\ket{n_\lambda}\!\bra{n_\lambda} \partial_\lambda \hat{H}_0(\lambda) \ket{m_\lambda}\!\bra{m_\lambda}}{E_n(\lambda)-E_m(\lambda)}.
\end{equation}
Then, in Eq.~\eqref{eq:coeffs}, only the diagonal part remains, and no transitions to other instantaneous eigenstates occur by construction. In the presence of near-degeneracies, care must be taken since the denominators may grow very fast. In a finite system, states that exactly cross belong to different symmetry sectors and then the sum in Eq.~\eqref{eq:cd-ham} must be disjointed into such sectors.

We can immediately picture the CD Hamiltonian for a two-level system with driven Hamiltonian
\begin{equation}
\hat{H}_0(\lambda) = \lambda \hat{\sigma}_z + \Omega \hat{\sigma}_x = \begin{pmatrix} 
\lambda & \Omega \\
\Omega & -\lambda
\end{pmatrix},
\label{eq:ham-LZ-1}
\end{equation}
where $\{\hat{\sigma}_x, \hat{\sigma}_y, \hat{\sigma}_z\}$ are the Pauli matrices. This gives
\begin{eqnarray}
\hat{H}_{\rm CD}(\lambda) = f_{\rm CD}(\lambda)\hat{\sigma}_y, \qquad f_{\rm CD}(\lambda)= -\frac{1}{2} \frac{ \dot \lambda \Omega }{\lambda^2+\Omega^2} \ .
\label{eq:cd-ham-2}
\end{eqnarray}
For a Landau--Zener (LZ) linear scan with $\lambda(t) \propto t$, the CD pulse $f_{\rm CD}(t)$ has Lorentzian shape in time.

Equation~\eqref{eq:cd-ham-2} shows what motivates the next section: $\hat{H}_{\rm CD}(\lambda)$ requires control of new operators (Pauli $\hat{\sigma}_y$), which were not present in $\hat{H}_0(\lambda)$ before. This turns out to be a recurring theme for generic adiabatic problems and, for certain classes of Hamiltonians, it can also be proven formally~\cite{Petiziol2018}. This issue makes the direct practical implementation of $\hat{H}_{\rm CD}(\lambda)$ challenging, motivating the search for approximate methods. 
 An intuitive explanation of why one generically expects $\hat{H}_{\rm CD}(\lambda)$ to contain operator components not included in $\hat{H}_0(\lambda)$ can be given as follows. The CD field can be understood as the generator of a continuous unitary transformation $\hat{U}_\lambda(\lambda)$ which diagonalizes $\hat{H}_0(\lambda)$, for each value of $\lambda$. Indeed, it can be rewritten as $\hat{H}_{\rm CD}(\lambda)=\dot\lambda \hat{\mathcal{A}}_\lambda$, where $\hat{\mathcal{A}}_{\lambda}=i\hbar \partial_\lambda \hat{U}_\lambda \hat{U}_\lambda^\dagger$ generates continuous translations with respect to $\lambda$, and has thus been termed \textit{adiabatic gauge potential}~\cite{Kolo2017}. A variation of the Hamiltonian can then be expressed as 
$\partial_\lambda \hat{H}_0(\lambda) = -\sum_n \partial_\lambda E_n(\lambda)\ket{n_\lambda}\! \bra{n_\lambda} + [\hat{\mathcal{A}}_\lambda, \hat{H}_0(\lambda)],$
which highlights that the variation of the instantaneous basis is ruled by $\hat{\mathcal{A}}_\lambda$. 
One can then verify that the adiabatic gauge potential is orthogonal (for instance, with respect to the Hilbert-Schmidt inner product) to both $\hat{H}_0(\lambda)$ and $\partial_\lambda \hat{H}_0(\lambda)$, and hence to the space spanned by those two operators. Unless the latter are a linear combination of a full set of operators spanning the whole operator space, and in the absence of particular symmetries, $\hat{H}_{\rm CD}(\lambda)$ will then typically contain new operators.

The natural question arises: if $\hat{H}_{\rm CD}(\lambda)$ cannot be implemented directly, can it still be approximated with the control parameters at hand, yielding an {\em effective} shortcut to adiabaticity? 
Ref.~\cite{Petiziol2018} answered this question positively using quantum control theory~\cite{DAlessandro:2007}, by showing that $-i\hat{H}_{\rm CD}(\lambda)$ belongs to the dynamical Lie algebra $\mathcal{L}$ of the controlled systems, for any value of $\lambda$. This formalizes the concept that the evolution generated by $-i\hat{H}_{\rm CD}(\lambda)$, for arbitrary $\lambda$, is always approximated arbitrarily well by a suitable time development of the control functions $c_i(t)$ of the control Hamiltonian 
\begin{equation}
\label{eq:ha}
\hat{H}(t) = \sum_{i=1}^N c_i(t) \hat{H}_i,
\end{equation}
where the operators $\hat{H}_i$ are those composing $\hat{H}_0(t)$, namely such that $\hat{H}_0(t)\in\mathrm{span}(\{\hat{H}_i\}_{i=1,\dots,N})$ for all times. $\mathcal{L}$ is defined as the smallest algebra spanned by $ \{-i \hat{H}_i\}_{i=1,\dots,N}$ and all possible nested commutators $[-i \hat{H}_l, [\dots,[-i \hat{H}_j, -i \hat{H}_k]]\dots ]$.

\section{Effective counterdiabatic driving}
\label{sec:eCD}

For the two-level system from above, an effective counterdiabatic (eCD) Hamiltonian, which approximates $\hat{H}_{\rm CD}(t)$ without controlling new operators, is given by
\begin{equation}
\label{eq:2-state-1}
\hat{H}_{\rm eCD}(t) = \sqrt{\frac{\omega}{2} \frac{\dot\lambda(t) \Omega} {\lambda (t)^2+\Omega^2}} \  [\sin(\omega t) \hat{\sigma}_z - \cos(\omega t) \hat{\sigma}_x].
\end{equation}
This Hamiltonian (assuming $\Omega>0$ and $\dot \lambda>0$ for all times) does indeed contain only the same operators as $\hat{H}_0(t)$ and hence avoids the use of new controls. It comes at the cost of adding sinusoidal drives, and it approximates the true $\hat{H}_{\rm CD}(t)$ well for large $\omega$. Note that, while $\hat{H}_0(t)$ of Eq.~\eqref{eq:ham-LZ-1} is purely real and thus invariant under time-reversal, the CD field of Eq.~\eqref{eq:cd-ham-2} is purely imaginary and breaks such a symmetry. The periodic forcing in Eq.~\eqref{eq:2-state-1} achieves this effectively through the anti-phase time dependence (co-sinusoidal and sinusoidal) of the $\hat{\sigma}_x$ and $\hat{\sigma}_z$ components, `circularly polarised' in the $x-z$ plane. Note also that the amplitude of the eCD field depends explicitly on $\lambda(t)$: eCD is not in competition with other methods to accelerate adiabatic protocols by optimising $\lambda(t)$~\cite{RolandCerf:2002, Rezakhani2009, Rezakhani2010}, but can be directly integrated with them for a combined speed-up.
In the following, we derive the form of the eCD Hamiltonian in Eq.~\eqref{eq:2-state-1} from a systematic procedure, based on matching the evolution operator generated by an \textit{ansatz} with the exact CD one, order by order in a Floquet-Magnus expansion for large $\omega$.

\textit{Effective CD construction.} The starting point for approximating $\hat{H}_{\rm CD}(t)$ is the Hamiltonian of Eq. \eqref{eq:ha}. An \textit{ansatz} is chosen to parametrise the control functions $c_i(t)$, whose free parameters are then fixed by enforcing that $\hat{U}(t)=\mathcal{T}\exp(-i\int_0^t \hat{H}(t') dt')$, where $\mathcal{T}$ denotes time ordering, approximates $\hat{U}_{\rm CD}(t)$ order by order in a perturbative expansion. The latter follows closely the paradigm of Floquet engineering~\cite{Goldman2014, Bukov2015, Eckardt2017}, though applied to the approximation of a time-dependent, rather than constant, Hamiltonian. We discretise the total protocol time $\tau$ into small steps $T$, within which the CD Hamiltonian $\hat{H}_{\rm CD}(t)$ is approximated as constant to order $T$. The CD field is discretised as $\hat{H}_{\rm CD}^{[n]} = \hat{H}_{\rm CD}(nT + T/2)$, with $n\in N$ and an error of order $O(T^3)$~\cite{Blanes2009}. The functions $c_i(t)$ are parametrised as time-periodic with angular frequency $\omega=2\pi/T$, using a truncated Fourier series with $L$ harmonics of $\omega$~\cite{Verdeny2014},
\begin{equation}\label{eq:contrfunc}
c_i(t) =  \sum_{j=1}^L \big[ A_{ij} \sin(j \omega t) + B_{ij} \cos(j \omega t)].
\end{equation}
The discretisation of time in steps of $T$ also defines sets of driving amplitudes $\{A_{ij}^{[n]}, B_{ij}^{[n]} \}$ pertaining to the $n$th time interval. Within each interval, we can apply Floquet theory~\cite{Eckardt2015}, thanks to the time periodicity of $c_i(t)$: the evolution operator admits a decomposition in the form $\hat{U}(t)=\hat{K}(t)e^{-i \hat{H}_F t}$, where $\hat{H}_F$ is the Floquet Hamiltonian governing the stroboscopic dynamics, $\hat{U}(T)=e^{-i \hat{H}_F T}$, while $\hat{K}(t)$ is a periodic operator, $\hat{K}(t+T)=\hat{K}(t)$, reducing to the identity at stroboscopic times $t = n T$. Hidden in the definition of $\hat{H}_F$ is a dependence on the initial time $t_0$ [or, equivalently, on the global phase of the oscillating functions in Eq.~\eqref{eq:contrfunc}], which we do not explicitly indicate, and the choice $t_0=0$ is assumed in the following. The role of driving phases will be discussed below. 

The Floquet Hamiltonian $\hat{H}_F$ can be computed systematically by means of a Magnus expansion in powers of $\omega^{-1}$~\cite{Bukov2015, Eckardt2017}. This has the form $\hat{H}_F=\sum_{n=1}^\infty \omega^{-(n-1)}\hat{H}_{F, n} $ with first terms reading~\cite{Eckardt2015, Mikami2016}
\begin{subequations} \label{eq:floquetmagnus}
\begin{align}
& \hat{H}_{F,1} = \overline{H}_{0},\\
& \hat{H}_{F,2} = \sum_{m=1}^{+\infty} \frac{1}{m}[\overline{H}_m, \overline{H}_{-m}] + \sum_{m\ne 0} [\overline{H}_0, \overline{H}_m] e^{i m \omega t_0}.
\end{align}
\end{subequations}
The quantities $\overline{H}_m$ represent Fourier components of $\hat{H}(t)$, namely $\overline{H}_m=(1/T)\int_0^T \hat{H}(t) e^{-im\omega t} dt$. We ask for the eCD and CD evolutions to coincide stroboscopically. For each time interval in the discretisation, we thus force $\hat{H}_F^{[n]}=\hat{H}_{\rm CD}^{[n]}$ by using the constant amplitudes $\{A^{[n]}_{ij},B^{[n]}_{ij}\}$ as free parameters. It is clear that the smaller the period $T$, the better the target evolution $\hat{U}_{\rm CD}(t)$ is sampled, and so in general $\hat{H}(t)$ must oscillate quickly with respect to the original time-dependence of $\hat{H}_{0}(t)$ and $\hat{H}_{\rm CD}(t)$. In this limit, one further expects that the amplitudes found in nearby intervals will be close in value, and thus an interpolation is possible yielding smooth controls $c_i(t)$. A single harmonic ($L=1$) can be sufficient to impose matching of the evolution to leading order in $\omega^{-1}$. Raising the number of harmonics $L$ in Eq.~\eqref{eq:contrfunc} introduces more free parameters that can be used to fulfil also the constraint equations produced by higher-order Magnus terms $\hat{H}_{F,n}$, improving the quality of the eCD approximation in a systematic manner. In many-body systems, such multi-tone drives also allow one to control complex many-body interactions~\cite{Petiziol2021, Petiziol2022}.

\begin{figure}[tb]
\includegraphics[width=\linewidth]{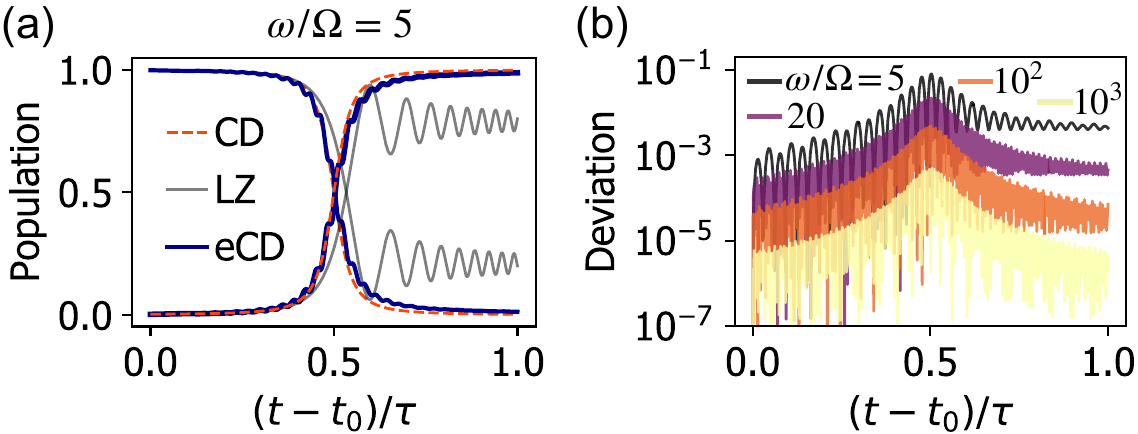}
\caption{(a) Example of evolution, of duration $\tau$ and starting at time $t_0$, driven by the eCD and adiabatic protocol for the LZ problem of Eqs.~\eqref{eq:ham-LZ-1} and~\eqref{eq:2-state-1}, of the populations of $\hat{\sigma}_z$ eigenstates. The eCD dynamics (dark blue) oscillates quickly but with small amplitude remaining close to the exact CD one, whereas the adiabatic LZ sweep (grey) fails to reach high fidelity with large amplitude oscillations. (b) Probability of nonadiabatic transitions for different driving frequency $\omega$. As $\omega$ is increased, both the stroboscopic and micromotion dynamics get closer and closer to exact CD.}
\label{fig:1}
\end{figure}

\textit{Two-level problem.} After having introduced the general method, the anticipated Eq.~\eqref{eq:2-state-1} for the LZ problem of Eq.~\eqref{eq:ham-LZ-1} is derived to illustrate the above ideas. Adopting a combination of the operators $\hat{\sigma}_x$ and $\hat{\sigma}_z$ that compose $\hat{H}_0(t)$, the eCD Hamiltonian is chosen of the form $\hat{H}(t) = c_x(t) \hat{\sigma}_x + c_z(t) \hat{\sigma}_z$.
To achieve compensation to first order in $T$ at stroboscopic times, we use a single harmonic ($L=1$) in Eq.~\eqref{eq:contrfunc}, and we choose the parametrisation 
\begin{equation}
c_x(t) = A \cos(\omega t + \phi); \quad c_z(t) = B  \sin(\omega t + \phi).
\end{equation}
A global phase $\phi$ has also been explicitly included to illustrate and discuss the role of driving phases below. The non-vanishing Fourier components of $\hat{H}(t)$ are 
\begin{equation}
\overline{H}_{\pm 1} = \frac{A}{2} e^{\pm i\phi}\hat{\sigma}_x \pm \frac{B}{2i} e^{\pm i\phi} \hat{\sigma}_z.
\end{equation}
Computing the Floquet Hamiltonian according to the Magnus expansion of Eqs.~\eqref{eq:floquetmagnus}, the first-order term, given by the time averaged Hamiltonian, vanishes ($\overline{H}_0=0$). The Floquet Hamiltonian is then determined, to leading order, by the second-order term and reads
\begin{equation}
\hat{H}_F = \frac{1}{\omega} [\overline{H}_1, \overline{H}_{-1}] =  \frac{A B}{\omega} \hat{\sigma}_y\ .
\end{equation}
Imposing the matching condition $\hat{H}_F^{[n]}=\hat{H}_{\rm CD}^{[n]}$ for the $n$th interval then gives the constraint equation $A^{[n]} B^{[n]}/\omega = f_{\rm CD}(n T + T/2) \ .$ We can replace here $\fb(t_n+T/2)$ by the continuous function $\fb(t)$ while maintaining the same level of approximation in $T$. A possible solution is then $A = \sqrt{\omega |\fb(t)|}, \quad B=\sign\big[\fb(t)\big] A$, where $\sign(x)$ indicates the sign of $x$ \cite{Petiziol2018}. In conclusion, an eCD Hamiltonian is obtained as 
\begin{equation}\label{eq:1stACD}
\hat{H}_{\rm eCD}(t) = \sqrt{\omega |\fb(t)| } \Big[ \sin(\omega t+\phi) \sz - \cos(\omega t+\phi) \sx\Big],
\end{equation} 
for $f_{\rm CD}(t)<0$ for all $t$, implying the anticipated Eq.~\eqref{eq:2-state-1}. A comparison between the adiabatic LZ evolution and that driven by the eCD field is shown in Fig.~\ref{fig:1}(a).

Note that the amplitudes found are proportional to $\sqrt{\omega}$. This occurs because $\hat{H}_{F}$ is determined by the second-order term in the Magnus expansion, where the amplitudes enter quadratically, and thus needs to be lifted to first order to match the CD field. 
This means that, if one increases the driving frequency to obtain a better sampling of $\hat{H}_{\rm CD}(t)$, then also the driving amplitudes must be scaled up. The impact of this on the performance of the eCD scheme as compared to the adiabatic approximation is analysed in detail in Ref.~\cite{Petiziol2018, Petiziol2019b}, confirming that eCD fields strongly enhance the protocol fidelity at given maximal driving power on timescales of practical interest.

\subsection{Micromotion and driving phase} With eCD, the effective Hamiltonian generated by the control fields approximates the CD Hamiltonian at stroboscopic times. What happens at intermediate times, in-between two full periods? Deviations from the dynamics generated by the Floquet Hamiltonian are described by the micromotion operator $\hat{K}(t)$. Analogously to the Floquet Hamiltonian, this can also be approximated through the Magnus expansion~\cite{Eckardt2015, Mikami2016}. It reads $\hat{K}(t)=e^{ \sum_{n>0} \omega^{-n} F_n(t,0)}$, where the leading order term $F_1(t,t')$ reads
\begin{align}
F_1(t,t') = -\sum_{m\ne 0} \frac{1}{m}(e^{im\omega t} - e^{im\omega {t'}})\overline{H}_m.
\end{align}
Already from the expansion of $\hat{K}(t)$, one can see that the micromotion reduces to the identity in the limit $\omega\to +\infty$. Therefore, as $\omega$ increases, not only the stroboscopic dynamics will match the adiabatic one closer, but deviations around the latter will also be suppressed for all times. This is illustrated in Fig.~\ref{fig:1}(b).

As mentioned above, the phase relation between $\hat{\sigma}_x$ and $\hat{\sigma}_z$ components in $\hat{H}(t)$ plays a crucial role, as it effectively reproduces the time-reversal symmetry breaking imposed by the exact $\hat{H}_{\rm CD}$. But what is the role of the global phase of the drives, and the impact of phase imperfections? As clear from the derivation of the eCD field for the two-level problem, the global phase $\phi$ of Eq.~\eqref{eq:1stACD} does not change the leading-order term in the Floquet Hamiltonian, $\hat{H}_{F,2}$, but it affects higher-order terms. Therefore, although for different choices of $\phi$ the target CD dynamics is always reproduced to leading order, the deviations due to higher-order terms exhibit a dependence on $\phi$. For some values of $\phi$, accidental error compensation may then also occur, yielding a fidelity increase. A relative phase offset between the $\hat{\sigma}_x$ and $\hat{\sigma}_z$ components has a more dramatic impact instead, since it also alters $\hat{H}_{F,2}$. A detailed analysis of such static-phase effects is found in Ref.~\cite{Petiziol2019a}.

\section{Applications: quantum state transfers}
We discuss now the application of the eCD scheme to specific problems, in particular for the realization of shortcut-to-adiabatic quantum state transfers and quantum gates.

\subsection{fmod-STIRAP}
\begin{figure}[t]
\includegraphics[width=\linewidth]{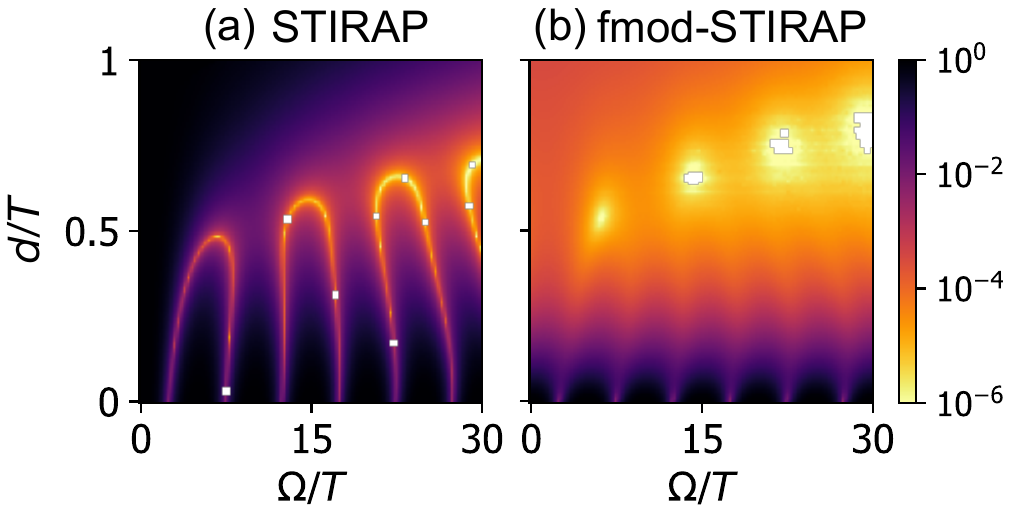}
\caption{Colour maps of the infidelity $1-|\langle 2|\psi_{\rm fin}\rangle|^2$, in log scale, for the end-of-protocol state $\ket{\psi_{\rm fin}}$ as a function of the Rabi frequency $\Omega$ and delay $d$ of Gaussian STIRAP pulses~\cite{Petiziol2020}. Panel (a) corresponds to STIRAP, while panel (b) corresponds to fmod-STIRAP. The evolution is computed over a time interval $t\in[-7\sigma, 3.5\sigma]$ and the eCD frequency used is $\omega=50\sigma$, where $\sigma/\sqrt{2}$ is the standard deviation of the STIRAP pulses. White points indicate unit fidelity within numerical precision.}
\label{fig:2}
\end{figure}

A first application of the effective CD method is in the context of STIRAP~\cite{Vitanov2017, Shore2017}---one of the most widespread adiabatic quantum control protocols. In its basic formulation, STIRAP enables population transfer between two states $\ket{0}$ and $\ket{2}$ which do not possess a direct matrix element, being it forbidden, for instance, by selection rules. This is achieved by separately coupling $\ket{0}$ and $\ket{2}$ to an intermediate state $\ket{1}$ with time-dependent pulses. A key feature of STIRAP is that it realises this desired population transfer without populating the intermediate state, thus not allowing leakage out of state $\ket{1}$. This is possible, since the STIRAP pulse sequence keeps the system in an instantaneous dark state at all times, which never overlaps with state $\ket{1}$~\cite{Vitanov2017}. The application of the exact counterdiabatic approach to STIRAP was the first example showing that the implementation of CD fields is typically challenging~\cite{Demirplak2003}: the CD pulse necessitates a direct coupling between the initial and target state, which by assumption is not possible. In rotating-wave approximation (RWA), the STIRAP Hamiltonian in two-photon resonance~\cite{Vitanov2017}, including the exact CD pulse $\Omega_{\mathrm{CD}}(t)$~\cite{Demirplak2003, GiannelliArimondo2014, VepsalainenParaoanu:2018}, reads
\begin{equation}
\hat{H}(t) = \frac{1}{2}\begin{bmatrix}
0 & \Omega_-(t) & i \Omega_{\mathrm{CD}}(t) \\
\Omega_-^*(t) & 2\Delta_1 & \Omega_+(t) \\
-i \Omega_{\mathrm{CD}}(t) & \Omega_+^*(t) & 0
\end{bmatrix},
\end{equation}
where $\Delta_1$ is the single-photon detuning. $\Omega_\pm(t)$ are typically referred to as pump ($-$) and Stokes ($+$) pulses. Following the eCD recipe, an eCD field approximating $\Omega_{\mathrm{CD}}(t)$ is constructed by adding new Fourier components to the pump and Stokes Rabi frequency at eCD frequency $\omega$, $\Omega_\pm(t)\to \Omega_\pm(t)+ \Omega_{\mathrm{eCD}}(t)\cos(\omega t + \phi_\pm) $~\cite{Petiziol2020}. Owing to the additional sidebands in the pulses, this protocol was dubbed frequency-modulated (fmod-) STIRAP~\cite{Petiziol2020}. As in the LZ case, the necessity to emulate a purely imaginary second-order coupling fixes the phase relation between the pulses to $\phi_+ - \phi_- = \pi/2$. Imposing that the oscillations reproduce the CD Rabi frequency then fixes the eCD amplitude to $\Omega_{\mathrm{eCD}}(t) = \sqrt{\omega \Omega_{\mathrm{CD}}(t)}$. A comparison of the final protocol (in)fidelity obtained by STIRAP and its accelerated eCD counterpart is shown in Fig.~\ref{fig:2} for different values of the Rabi frequency and delay $d$ between the Gaussian-shaped~\cite{Petiziol2020} pump and Stokes pulses. These results highlight how the eCD correction gives access to much larger fidelities, and in a much broader parameter regime, as confirmed by recent experiments \cite{Ge2023}.
Approximate CD fields for STIRAP implemented in a superconducting qubit~\cite{VepsalainenParaoanu:2019}, building on earlier proposals~\cite{GiannelliArimondo2014, ChenMuga2010, VepsalainenParaoanu:2018} can be understood as an eCD field similar to fmod-STIRAP, but using single-exponential (single sideband), rather than (co)sinusoidal (symmetric sidebands), additional pulses. A consequence of this is the generation of additional effective light shifts that must be compensated for, which would instead be cancelled by construction in fmod-STIRAP.

\begin{figure}[t]
\centering
\includegraphics[width=0.7\linewidth]{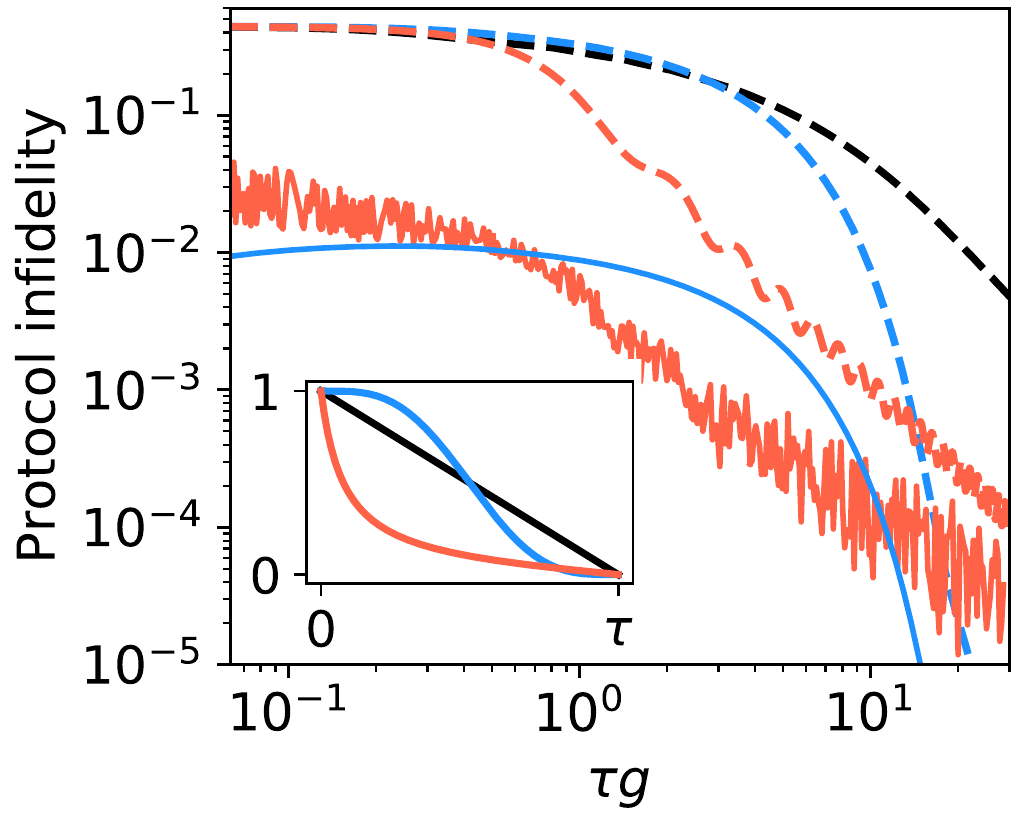}
\caption{End-of-protocol infidelity vs. the duration $\tau$ for the preparation of a Bell state in circuit QED. The curves correspond to the adiabatic (dashed) and eCD (solid) cases, with the different colours corresponding to the different ramp functions of same colour in the inset. Besides the linear LZ ramp (black), the ramps include a locally-adiabatic sweep~\cite{Petiziol2019b, Rezakhani2009} (red) and a polynomial boundary-cancellation method~\cite{Petiziol2019b, Rezakhani2010} (blue).}
\label{fig:3}
\end{figure}

{\itshape Bell states in circuit quantum electrodynamics (QED).} A second example is the preparation of entangled Bell states with two qubits dispersively coupled to a photonic mode, acting as a quantum bus~\cite{Petiziol2019b}. This setup describes, for instance, a circuit QED system of transmon qubits coupled to a common microwave resonator~\cite{Majer2007, Krantz2019, Blais2021}. The system Hamiltonian can be written in the form $\hat H = \hat{H}_Q + \hat{H}_{QR} + \hat{H}_R$, where $\hat{H}_Q =\sum_{i=1,2} \Omega_i \hat{\sigma}_i^z/2$ and $\hat{H}_R=\omega_R\hat{a}^\dagger\hat{a}$ are the bare qubit and resonator Hamiltonians respectively, while $\hat{H}_{QR}=\sum_{i=1,2} g_i (\hat{\sigma}_i^+\hat{a} + \hat{\sigma}_i^-\hat{a}^\dagger)$ describes the qubit-resonator coupling in RWA. Here, $\hat{a}^\dagger$ and $\hat{a}$ represent creation and annihilation operators for photons in the resonator, respectively. When the qubits are close to resonance with each other ($\Omega_1\approx \Omega_2$), in the dispersive qubit-resonator regime ($|\Omega_i - \omega_R|\gg g$), the resonator mediates a second-order effective coupling between the qubits of strength $\tilde{g} = g_1 g_2/[2(\Omega_1 - \omega_r)]$, describing a flip-flop interaction $ \tilde{g}({\hat{\sigma}}_1^+ {\hat{\sigma}}_2^- + {\hat{\sigma}}_1^- {\hat{\sigma}}_2^+ )$. The single-excitation eigenstates at resonance are thus entangled Bell states $\ket{B_\pm}=(\ket{01}\pm\ket{10})/\sqrt{2}$. 
In Ref.~\cite{Petiziol2019b}, an adiabatic preparation protocol of such states is developed. An initial offset in the qubit frequencies is ramped down adiabatically until resonance is reached, where an avoided crossing of width $\sim2\tilde{g}$ signals the second-order coupling. Assuming $\Omega_1>\Omega_2$, the adiabatic evolution maps $\ket{01}\to\ket{B_+}$ and $\ket{10}\to\ket{B_-}$ up to phase factors. However, the second-order nature of the coupling requires long ramping times to avoid excitations, and the application of a shortcut is thus particularly instrumental. When computing numerically the exact counterdiabatic field for this protocol, its major component is a time-dependent interaction $\hat{H}_{\rm CD}(t) = i h(t)(\hat{\sigma}_1^+ \hat{\sigma}_2^- - \hat{\sigma}_1^- \hat{\sigma}_2^+)$. An eCD field for this problem is constructed in Ref.~\cite{Petiziol2019b} by considering time-dependent modulations of the qubit-resonator couplings. Such modulations are directly possible in tuneable-coupling superconducting qubits~\cite{Gambetta2011}, or can be interpreted as the effect of modulations of the resonator-qubit detuning (i.e. of the qubits' or resonator's transition frequencies). The eCD correction takes the form
\begin{equation}
\hat{H}_{\rm eCD}(t) = \sqrt{2\omega h(t)}[\cos(\omega t) \hat{\sigma}_1^+ +\sin(\omega t) \hat{\sigma}_2^+] \hat{a} + \mathrm{H.c.}
\end{equation}
The performance, in combination with different choices for the adiabatic ramp function, shows a substantial speed-up of the protocol at desired fidelity values \cite{Petiziol2019b}. This is exemplified in Fig.~\ref{fig:3} for a selection of sweep functions. By combining an optimised ramp with eCD, the protocol infidelity can be improved by orders of magnitude in comparison to a simple linear ramp.

\section{(Shortcut-to-)Adiabatic quantum gates} Going beyond individual state transfers, eCD fields are used to speed up full quantum gates building on adiabatic evolution, thus finding immediate applications in quantum information processing. Gate synthesis is in general a more challenging task, since it requires implementing a whole unitary, rather than controlling the evolution for one specific initial state only. We briefly discuss  two examples, where the eCD framework is used to speed up an arbitrary single-qubit gate based on a modified STIRAP protocol and a geometric controlled-phase gate in Rydberg atoms. 

\subsection{fSTIRAP-based single-qubit gate} The first example is the acceleration of an arbitrary adiabatic single-qubit gate based on a modified version of STIRAP, known as fractional STIRAP (fSTIRAP)~\cite{VitanovShore1999}. The latter implements the STIRAP pulse sequence with modified pump and Stokes pulses, which depend on two new parameters $\eta$ and $\chi$ that control the final superposition state obtained, $\psi_{\mathrm{fin}}(\eta, \chi) = \cos(\eta)\ket{0} - e^{-i\chi}\sin(\eta)\ket{2}$. By implementing a sequence of two non-resonant ($\Delta_1\ne 0$) fSTIRAPs, one with pulse order swapped, a complete gate 
\begin{equation}
\hat{U}(\eta, \chi)=\begin{bmatrix}
\cos(2\eta) & e^{i\chi}\sin(2\eta)\\
 -e^{-i\chi} \sin(2\theta)& \cos(2\theta),
 \end{bmatrix}
\end{equation}
is obtained. A double fSTIRAP is needed, since a single one would introduce an unwanted phase spoiling the gate. This adiabatic gate can be accelerated by using an eCD correction identical in matrix form to the fmod-STIRAP case, for $\chi=0$, but involving a Rabi frequency adapted to the new shape of the pump and Stokes pulses pulses. Details on the exact shape of the eCD pulse and a performance study are found in Ref.~\cite{Petiziol2020}.

{\itshape Geometric two-qubit gate from Rydberg blockade.} In the context of Rydberg atoms~\cite{Gallagher2005, Saffman2010, Morgado2021}, a controlled-phase two-qubit gate between neighbouring atoms can be realised, based on the accumulation of Berry phase over a cyclic operation, by exploiting the Rydberg blockade effect~\cite{Saffman2020}. To reach high fidelities, strong Rabi frequencies would be desirable to enhance adiabaticity, but, at the same time, they increase the value of an additional dynamical phase imprinted to the state, which scales roughly as the ratio between Rabi frequency and Rydberg blockade. The latter phase is unwanted, and represents a gate imperfection, such that a compromise between this error and nonadiabatic errors must be accepted at finite $V$. Both issues can be mitigated by eCD \cite{Luis2023}. The eCD field, besides counteracting nonadiabatic effects, is able to bypass the creation of the unwanted phase by approximating $\hat{H}_{\rm CD}(t)$ without implementing also the original adiabatic pulses, which are responsible for the dynamical phase. Moreover, Ref.~\cite{Luis2023} proposes an approximation for a separable pulse to overcome the problem that the CD pulses typically require control on additional qubit-qubit interactions which are hard to implement.

\section{Conclusion}
\label{sec:concl}
We reviewed the construction of effective CD fields for accelerating quantum control protocols based on adiabatic driving. We discussed some applications spanning from the realisation of high-fidelity state transfers to the design of full quantum gates. A natural outlook is to extend these methods towards STAs for truly many-body quantum systems. For large systems, the problem of approximating the CD field is further complicated by the fact that the exact CD field cannot be determined neither analytically nor numerically, when exact diagonalisation of $\hat{H}_0(t)$ is over-demanding~\cite{Kolo2017, Sels2017}. A variational method to derive approximate CD fields without knowledge of the spectral properties has been developed in~\cite{Sels2017}, and used to propose a systematic method to Floquet-engineer STAs~\cite{Claeys2019}, similar in spirit to the approach reviewed here. An interesting perspective is to combine the latter methods in order to broaden the range of STA protocols for quantum many-body dynamics. 

\section*{Acknowledgments}
We thank E. Arimondo, S. Carretta, B. Dive, L. Giannelli, R. Mannella, and L. Yag\"ue Bosch for our fruitful collaboration.
Supported by the National Recovery and Resilience Plan, Mission 4 Component 2 Investment 1.3 -- Call for tender No. 341 of 15/03/2022 of Italian MUR funded by NextGenerationEU, with project No. PE0000023, Concession Decree No. 1564 of 11/10/2022 adopted by MUR, CUP D93C22000940001, Project title ``National Quantum Science and Technology Institute".


apsrev4-2.bst 2019-01-14 (MD) hand-edited version of apsrev4-1.bst

\end{document}